# Quantum Size Effect transition in percolating nanocomposite films

**B. Raquet*[1], M. Goiran[2], N. Nègre[1], J. Léotin[1]**

[1] Laboratoire de Physique de la Matière Condensée de Toulouse, LPMCT-LNCMP-INSA, Av. de Rangueil, 31077 Toulouse, France.
[2] Laboratoire National des Champs Magnétiques Pulsés, Av. de Rangueil, 31077 Toulouse, France.

**B. Aronzon[1,2], V. Rylkov[1,2], E. Meilikhov[1]**

[1] Russian Research Center "Kurchatov Institute", 123182 Moscow, Russia
[2] Scientific Center for Applied Problems in Electrodynamics, Izhorskaya 13/19, Moscow 127412, Russia.

**Abstract** : We report on unique electronic properties in $Fe-SiO_2$ nanocomposite thin films in the vicinity of the percolation threshold. The electronic transport is dominated by quantum corrections to the metallic conduction of the Infinite Cluster (IC). At low temperature, mesoscopic effects revealed on the conductivity, Hall effect experiments and low frequency electrical noise (random telegraph noise and *1/f* noise) strongly support the existence of a temperature-induced Quantum Size Effect (QSE) transition in the metallic conduction path. Below a critical temperature related to the geometrical constriction sizes of the IC, the electronic conductivity is mainly governed by active tunnel conductance across barriers in the metallic network. The high *1/f* noise level and the random telegraph noise are consistently explained by random potential modulation of the barriers transmittance due to local Coulomb charges. Our results provide evidence that a lowering of the temperature is somehow equivalent to a decrease of the metal fraction in the vicinity of the percolation limit.

**Introduction**

The results of early studies on electrical properties of Metal-Insulator composites or metal granular films were successfully considered in the framework of percolation and effective medium theories.[1-3] When the metal volume fraction $x$ is less than percolation threshold $x_c$, the conductivity occurs via inter-granular tunnelling or temperature activated hopping. Its temperature dependence follows the well known $1/\sqrt{T}$ law, i.e. $\sigma(T) \propto exp(-\sqrt{T_0/T})$.[4-6] As the system approaches the percolation limit, the correlation length $L_c$ diverges and the conductivity is mainly governed by the physical properties of the Infinite Cluster (IC) formed by metallic grains in contact. For $1-x \ll 1$, the granular film behaves like a macroscopically homogeneous disordered metal and the conductivity at low enough temperatures is mainly governed by quantum corrections due to Weak Localisation (WL)[7] and renormalization of the Electron–Electron Interaction (EEI)[8], both revealed in several measurements.[9-11]

Recently, the interest for magnetic and granular materials has been reinforced because of unique properties, which are much more pronounced in the close vicinity of the percolation threshold. Let us mention, for instance, the giant magnetoresistance (GMR)[12-15], the giant extraordinary Hall effect[16,17], nonlinear transport[18] and optical properties[19] widely discussed in the literature. However the exact mechanisms of these effects remain elusive. Even more, the physical origin of the famous $1/\sqrt{T}$ law for the temperature dependence of the conductivity is controversially discussed within different theoretical concepts.[5,6,20] Discrepancies between theories and experimental results mainly originate from various quantum effects such as single-electron transport effects[21], quantum interference[7,8] and quantum size effects, which strongly affect the granular metal properties[9-11,13,14,22] and are not properly described by the classic theories of the electronic transport.

The present paper is devoted to the study of Quantum Size Effects (QSE) on the electronic transport in Fe$_{X \approx Xc}$-SiO$_2$ nanocomposite (granular metal) films, in the proximity of the percolation threshold. We performed magneto transport (both magneto resistance and Hall effect) and electronic noise measurements to probe the static and dynamical electronic properties of a set of percolating channels. Hall effect and resistance fluctuations undergo an unexpected temperature transition at low temperatures. Our results provide evidence for a temperature-induced QSE transition in the Infinite Cluster (IC) resulting in a crossover from tunnelling to metal-type conduction. In the vicinity of the temperature-induced transition, we report on an astonishing Random Telegraph Noise (RTN) which reveals temperature-induced changes in the IC conductivity. To our knowledge, it's the first observation of RTN in macroscopic granular films. We use a frequency-temperature analysis of the non-gaussian noise to gain insight into the microscopic origin of the resistance fluctuations. A 100 meV activation energy for the fluctuating process has been inferred and the RTN is interpreted in term of drastic current redistribution in the overall connectivity of the metallic network. We attribute the temperature-induced transition in IC to thermally activated cut off in the percolating path induced by Quantum Size Effects. The magnetic contribution to the electronic transport in the Infinite Cluster is also discussed.

**Experiment**

$Fe_x$-$SiO_2$ nanocomposite films are prepared by the ion-beam sputtering technique in a vacuum chamber with a mosaic target consisting of Fe and $SiO_2$ tablets. The volume fraction of iron *x* varies between 0.3 and 0.9 and is controlled by *X*-ray micro-analysis. The film thickness *t* are about 0.4 µm. Transmission electronic microscopic (TEM) has been performed to estimate the iron grain size distribution. The distribution function *f(L)* of grain sizes for a $Fe_x(SiO_2)_{1-x}$ film similar to those used for electrical measurements is shown in Fig.1, with a Gaussian fit in solid line and an average grain size value around 4 nm. The distribution ranges from about 8 nm down to isolated atoms. Such a widespread distribution toward the atomic scale is likely inherent to composites prepared by sputtering.[23] The resistivity dependence on the metal content exhibits a percolation transition at $x_c \approx 0.56$. The rather high value of the percolation threshold is due to the significant amount of iron in atomic state. The results we present below focus on Fe$_x$-SiO$_2$ samples in the very close upper percolation threshold, $x \approx x_c^+$. In the hopping regime, magnetic and electronic properties have been presented before.[15] The magneto transport measurements, up to 40 T, are performed on patterned samples into a Hall bar geometry of 7 mm length, 2 mm width and a 2.5 mm distance between potential probes. Electrical noise measurements are based on the standard 4 probes DC technique from 4.2 to 300 K.[24] Special care was taken to ensure that spurious noise like contact noise does not contribute to our results. The experimental background noise is in agreement with the theoretical Johnson noise of the sample and we checked the square current dependence of the excess noise spectral density, $S_v(f) \propto I^2$. 1/f noise is measured with a current density lower than $10^2$ A/cm$^2$ to prevent heating effects. The noise data we present correspond to the power spectral density of the voltage normalized to the applied current and the resistance of the sample.

**Results and Analysis**

Fig. 2 presents the temperature dependence of the conductivity of Fe$_{X-Xc}$-SiO$_2$ granular films representative of a panel of results obtained in the very close upper vicinity of the percolation threshold. The conductivity continuously increases from 1.8 K to room temperature by a factor of 3. This rather weak temperature dependence for a granular system and the non-exponential behaviour are consistent with a metallic conductivity regime above the percolation threshold.[2,10,12,25] Various laws of conductivity variations with temperature have been suggested in the literature, ranging from the logarithmic to a power law behaviour $\Delta\sigma(T) \propto T^\beta$, $\beta<1$, to deal with disordered metallic systems and nanocomposites in the metallic regime.[7-11] They mainly reflect quantum corrections at sufficiently low temperature to the conductivity process which originates from WL and renormalization of the EEI in a random potential. In first approximation, both processes are additive and in a 3D system, the conductivity corrections vary as follow: $\Delta\sigma(T) \approx A_{WL}T^{p/2} + A_{e-e}\sqrt{T}$.[7] $A_{WL}$ and $A_{e-e}$ are constants related respectively to the WL and EEI. The *p* exponent varies from 3/2 to 3 depending on the scattering mechanism. In 2D systems, both processes exhibit a logarithm dependence. Our conductivity data are well described by a $\sqrt{T}$ law in a wide range of temperature, from 300K to 70K. Such signature of E-E Interaction at rather high temperature is not unusual for granular metallic films[10-11,25]; the strength of the EEI corrections being directly coupled to an increased disorder.

Below 70K, a significant departure from the square root law is observed with a much-pronounced increase of the resistivity. In any case, our results do not follow a ln(*T*) dependence of the conductivity, even at low temperatures. It rules out any dimensionality change in the characteristic conductivity lengths, but it is a hint of a temperature-induced transition toward a more resistive regime.[10] To go through details of the electronic transport in the Infinite Cluster, we focus on Hall effect, noise measurements and mesoscopic effects on the conductivity.

In Fig. 3 are presented the magnetic field dependencies of the Hall resistance $\rho_{Hall}(H)$ for several temperatures. All the curves exhibit two well-defined regimes corresponding respectively to the extraordinary ($R_e$) and ordinary ($R_o$) Hall effect. In low fields, the sharp increase of the Hall voltage is related to the magnetization through the strength of the spin-orbit coupling between the charge carriers and the magnetic moments. In the high field regime, the Hall voltage follows a linear and much weaker field dependence.[26] The high magnetic field measurements of the Hall effect provide an accurate estimate of $R_0 = \partial\rho_{Hall}(H)/\partial H\big|_{H\gg H_s}$ the ordinary Hall coefficient, well above the technical saturation of the magnetization. Its temperature dependence is presented in the inset of Fig. 3. An astonishing variation of $R_0$ is observed around 70 K. Above and below of this "critical" value, the Hall voltage remains practically temperature independent. From the percolation theory[1] and the $R_0$ values, we infer a rough estimate of the effective charge carriers density involved in the percolating channels. At room temperature, $n_{eff} \approx 9.10^{22}$ cm$^{-3}$. This is comparable to the bulk metal value, which is consistent with granular systems with a metallic fraction above the percolation threshold. Below 70 K, the system behaves as if the number of carriers participating to the conduction is drastically reduced by one order of magnitude. The abrupt rise of $R_0$ reinforces the assumption of a temperature-induced transition at $T_t \approx 70$ K.

The geometrical misalignment $d_{Hp}$ of the Hall contacts in a patterned sample is one of the sources of non-zero Hall voltage $V_{Hp}$ in zero-magnetic field. In an homogeneous sample, $V_{Hp}$ is equal to $RId_{Hp}/d$, where $R$ and $d$ are respectively the resistance and distance between the resistance potential probes and $I$ is the applied current. In our samples the geometrical misfit is estimated to be about 5 µm. However, it is well established that additionally to the pure misfit voltage, intrinsic inhomogeneities in the IC at a scale less than the correlation length, originating from the IC random nature, strongly contribute to $V_{Hp}$.[27] So, in a general case, $V_{Hp}(T)$ can be rewritten as follows : $V_{Hp}(T) = RId_{Hp}/d + \Delta V_{Hp}(T)$, where $\Delta V_{Hp}(T)$ is the local voltage arising in the different percolation paths probed in-between the two Hall contacts. From the above expression and assuming that, at room temperature $V_{Hp}$ originates from the geometrical probes misfits only, we can extract the voltage $\Delta V_{Hp}(T)$ which is of incoherent mesoscopic nature[27]. The Fig. 4 represents the temperature variation of $\Delta V_{Hp}(T)$ obtained for one of the Hall probes location on the sample. The drastic temperature dependence which significantly differs from the one of the overall resistivity of the sample provides an evidence for IC realignment. The local voltage $\Delta V_{Hp}(T)$ can be expressed as a function of the correlation length $L_c$ in the metallic network : $\Delta V_{Hp} \approx 2E \, L_c$[27], where $E$ is the averaged electrical field over the sample. Therefore, the temperature dependence of $\Delta V_{Hp}(T)$ can be related to the temperature-induced changes of the correlation length in the Infinite Cluster (inset Fig. 4). $\delta L_c$ clearly peaks at $T = T_t$ with a 1 µm estimated $L_c$ peak value assuming a negligible correlation length at room temperature. This striking mesoscopic effect strongly supports the assumption of a temperature-induced percolation transition in the metallic path as $L_c$ theoretically diverges at the percolation threshold. The results obtained on distinct locations of the Hall pairs provide the similar $L_c(T)$ dependencies.

To probe the microscopic origin of the conductivity and its temperature-induced transition, we measure the low frequency electronic noise between $10^{-2}$ to $10^2$ Hz. The power spectral density of the noise $S_V(f)$ mostly follows a $1/f^\alpha$ dependence from 4 to 300 K, with α close to 1 (inset Fig. 5). The noise level versus temperature at 10 Hz reveals a sharp variation by two orders of magnitude around 70 K (Fig. 5). This spectacular behavior of the electronic noise occurs at the same temperature $T_t$ as the abrupt variation of the ordinary Hall effect and the departure of the conductivity toward a more resistive regime. Below and above the jump, the noise levels are almost constant; only a slight increase of the noise is noticeable above 130 K. The excess noise is commonly defined by the empirical Hooge's expression[28]: $S_v(f) = v^2 \gamma / N f^\alpha$, where γ is the Hooge's constant, $v$ the applied voltage and $N$ the charge carriers number in the noisy volume. From the $S_v(T,f)$ values and the high field Hall effect, we deduce $\gamma \approx 300$ at 300 K. This noise level is at least 4 orders of magnitude greater than normally found in well crystallized metallic thin films.[29] Nevertheless, it remains rather small compared to the γ values usually found in granular systems dominated by the hopping regime.[30] Below 70 K, the Hooge's constant is significantly increased and reaches 2000.

In the close vicinity of the transition, between 50 and 65 K, the resistance fluctuations are dominated by a large and unexpected non gaussian noise (Fig. 6a). Random telegraph noise is observed with resistance steps of the order of $\Delta R/R \approx 4 \cdot 10^{-7}$ at 50 K. The RTN has been theoretically predicted in percolating macroscopic systems with an intrinsic sporadic behavior over time.[31] In our case, the discrete switching events are sometimes barely resolvable above the background gaussian fluctuations which prevents any statistical approach in the time domain. Nevertheless, the power spectral density of the time traces reveals thermally activated features related to the non gaussian noise. The $S_v(f,T) \times f$ spectra versus frequency show bumps whose frequency peaks $f_p$ shift with the temperature. These bumps are a clear signature of a well-defined fluctuating process responsible for the RTN.[32] The frequency-temperature dependence of the bumps follows an Arrhenius law: $f_p = f_0 \exp(-E_a/kT)$. $E_a$ is the activated energy of the fluctuating process and $f_0$, the attempt frequency in a two levels model. The fit in inset Fig. 6b yields $E_a \approx 100$ meV and $f_0 \approx 2.8 \cdot 10^{-10}$ s$^{-1}$. The attempt frequency is consistent with a phonon assisted fluctuating process. It is worth mentioning that the Dutta-Dimon-Horn model[33] mainly used to deal with noise in metallic films is in obvious disagreement with our results around $T_t$, even if one involves the temperature dependence of the charge carriers number given by the Hall voltage data. We conclude that drastic changes occur in the nature of the fluctuators and their coupling to the conductivity in the vicinity of $T_t$. Fluctuating mechanisms with an activated energy centered on 100 meV widely dominate the dynamic process around $T_t$.

In addition, we performed SQUID measurements, high magnetic field Kerr effect and magneto resistance to probe the magnetization and its influence on the conductivity. The temperature dependence of the conductivity doesn't exhibit significant changes as a function of an applied magnetic field up to 30 T. The Kerr data and the ZFC-FC magnetization obtained by SQUID, show the coexistence of two magnetic phases, one is ferromagnetic and persists up to room temperature, the other one is superparamagnetic with a blocking temperature distributed around 90 K.

**Discussion**

Our results reveal the existence of a peculiar temperature-induced electronic transition in the IC for which we concomitantly observe a sharp increase of the ordinary Hall coefficient, a peak of the correlation length and an enhancement of both the resistivity and its fluctuations as well as a RTN. These observations strongly suggest an abrupt realignment of IC, i.e. changes in the percolating network, with a reduction of its effective volume.

In an attempt to understand the physical origin of the IC breaks or realignment, it is of interest to deal with the particular structure of the percolating paths as the metallic fraction tends to $x_c$. The main feature of the conducting channels is the existence of bottlenecks constituting conduction bridges between larger metallic parts. If we assume that those weak links correspond to geometrical constrictions of iron particles in contact, they should represent a privileged location for Quantum Size Effects (QSE) induced by the electronic confinement. QSE result in the electronic level splitting within the constriction which induces an energy barrier $\Delta$ for conduction electrons. Such geometrical constrictions have been observed in our sample by electronic microscopy with an average size ranging up to approximately 3 nm.

In the simplest QSE model, the height of the barrier is approximated by $\Delta = 1/[N(\varepsilon_F) L^3] \sim \varepsilon_F/(k_F L)^3$ where $N(\varepsilon_F) \propto m^{3/2} \varepsilon_F^{1/2}/h^3$ is the electronic density of state at the Fermi level and $L$ is the size of a small granule connecting two granules of larger sizes.[38] The above expression yields 10 meV energy barrier for a 3 nm constriction.

At first, let us consider the conductivity $\sigma(T)$ at temperatures above 70K. In this temperature range, $\sigma(T)$ is strongly affected by corrections due to the E-E Interaction. The conductivity variations with temperature caused by the corrections $\Delta\sigma$ is of the order of its whole value, $\Delta\sigma/\sigma$ is roughly equal to unity. On the contrary, the Ioffe and Regel criteria yields $\hbar/(\tau\varepsilon_F) \leq 3\ 10^{-2} \ll 1$, where $\varepsilon_F$ and $\tau$ are respectively the Fermi energy and the electron momentum relaxation time inside grains. As a matter of fact this apparent contradiction comes from the strongly enhanced quantum corrections in the close vicinity of the percolation transition (see, for instance, Ref. [9, 34]). Even more, the percolation threshold proximity itself provokes Anderson localisation in weak-links or bottlenecks of the percolation cluster.[35,36] Let us define $\varepsilon_\Delta$ as the electron kinetic energy over the top of the barrier induced by QSE in constrictions. The electron overcoming the barrier has a wavelength $\lambda_\Delta \sim h/(2m\varepsilon_\Delta)^{1/2}$, much longer than the value $\lambda_F \sim h/(2m\varepsilon_F)^{1/2}$, characteristic for the IC "lakes". For small enough $\varepsilon_\Delta$-values, i.e. $\lambda_\Delta \gg l \sim L$, where $l$ is the electron mean free path, electrons in the constriction are localised. This mechanism gives rise to Anderson localisation near the percolation threshold and was observed in $Pd_xC_{1-x}$ granular metallic films[10] with structure similar to our system.

Now let turn to the discussion of the transition. Below 70 K, the Hall effect demonstrates an anomalous reduction of the effective volume of conductivity. The direct consequence of those local breaks in the metallic network is the reduction by one order of magnitude of the available number of carriers. It is of interest to notice that the temperature dependence of the ordinary Hall effect is in qualitative agreement with the Hall effect variation versus the metallic fraction near $x_c$ predicted by Shklovskii.[39] This model, based on a scaling hypothesis with two different electrical conductivities constituting the system, gives rise to a sharp increase of the ordinary Hall coefficient as the metallic fraction is decreased down to $x_c$. Therefore, in the close vicinity of the percolation, a decreasing of the temperature affects the conductivity in the Infinite Cluster in the same way as a small reduction of the metallic fraction of iron near $x_c$. Such an analogy is fully consistent with the temperature dependence of the correlation length $L_c$ (inset Fig. 4).

However, the abrupt change of the Hall effect compared to the smooth one of the conductivity near $T_t$ cannot be explained in the frame of a single mechanism of conductivity. Following the Shklovskii model[39], we analyse the conductivity and the Hall effect as a result of two distinct conductances $G_{IC}(T)$ and $G_{hop}(T)$ contributing to the electronic transport. $G_{IC}(T)$ is the conductance through the IC and $G_{hop}(T)$ corresponds to the hopping regime over isolated granules, $G_{hop}(T) \propto exp\left(-\sqrt{T_0/T}\right)$. From our conductivity data, we deduce the temperature dependence of the two conductances : the $T_0$ value is estimated around 12 K and $G_{IC}(T)$ is inferred from the [$G(T)$- $G_{hop}(T)$] difference (Fig. 7). The rather small $T_0$ value is in agreement with granular films in the hopping regime close to the percolation. The temperature dependence of $G_{IC}(T)$ (inset Fig. 7) reveals a drastic crossover at $T \approx 60$ K, close to $T_t$. At low temperatures, $G_{IC}(T)$ is practically temperature independent and dominates the overall conductivity. Well above 60 K, $G_{IC}(T)$ roughly follows the $\sqrt{T}$ law in accordance with quantum corrections to the metallic conductivity. Finally, the IC conductivity in first approximation can be represented as $G_{IC}=G_0$ at $T<T_t$ and $G_{IC} \propto \sqrt{T}$ at higher temperatures.

We interpret the occurrence of the two clearly defined conductivity regimes by the predominant role of the QSE at low temperature ( $T< T_t$ ) leading to sizeable energy barriers in the smallest granules which contact clusters of bigger size. For $T>T_t \approx \Delta$ the barriers are smeared out by the temperature, the conductivity is metallic, while at low temperatures, electrons can tunnel through the barrier and the conductivity is temperature independent like the barrier transmittance.

At a given temperature, the conductance $g(T)$ of the tunnel constriction is defined by thermal activated tunnel transitions[41]:

$$g(T) \propto exp(-L/\lambda_\Delta)(\pi kT/\varepsilon_L)/ sin(\pi kT/\varepsilon_L)$$

where $\lambda_\Delta \sim h/(2m\Delta)^{1/2}$, and $\varepsilon_L=\Delta(\lambda_\Delta/L) \sim h^2/mL^2(k_F L)^{1/2}$. The cited relation is valid at $kT/\varepsilon_L<1$.

At low temperatures the constriction conductance is minimum and can be considered as a "switched off" one. However, starting from the temperature $T \sim T_L=\varepsilon_L/k$, the conductivity properties of the constriction begin to increase and its conductance rises. Formally, $g \to \infty$ at $T \geq T_L$, so the energy $\varepsilon_L \propto L^{-5/2}$ corresponds to an activation energy for the constriction of size $L$. It means that, at a given temperature, only the contacts with $L>L_T$ where $L_T \sim k_F^{-1}(\varepsilon_F/kT)^{2/5}$ are "switched on". All constrictions of smaller sizes are switched "off" at this temperature.

We argue that the temperature dependence of the IC conductivity is strongly affected by the above mechanism of constrictions' "switching on" under a temperature increase. In the "open" (switched on) constriction, the electron wave length $\lambda_\Delta$ is long enough to satisfy the inequality $\lambda_\Delta > L$. Thus, the conductance of such a constriction is defined by a weak electron localization and depends on temperature as follows: $g = g_0 + Const \cdot T^{1/2}$ where the parameter $g_0$ is determined by the condition $g(T_L)=0$, that is $g(T) \propto (T^{1/2} - T_L^{1/2})$.

The temperature dependent part $G_T$ of the IC conductivity is straightforward expressed as the sum of conductances of "open" constrictions:

$$G_T \propto \int_0^T f(T_L)\left(\sqrt{T} - \sqrt{T_L}\right) dT_L .$$

Therefore, we can write :

$$f(\varepsilon_L) \propto \partial\left[\sqrt{T}(\partial G_T / \partial T)\right]/\partial T \Big|_{T=\varepsilon_L/k} .$$

Hence, if we extract the relevant part of the conductance $G_T$, the distribution $f(\varepsilon_L)$ of constrictions over the activation energies can be inferred as shown in Fig. 8 (small circles). The noisy aspect of the curve is inherent to a second derivative of experimental data. The distribution is finally approximated by a log-norm function (solid line in Fig. 8) :

$$f(\varepsilon_L) \propto exp\left(-Ln^2(\varepsilon_L / \varepsilon_0)/2\sigma^2\right),$$

with $\varepsilon_0 = 1.35$ meV and $\sigma = 1$. It means that the average constriction activation energy is equal to $\langle \varepsilon_L \rangle = \varepsilon_0 exp(3\sigma^2/2) = 6$ meV, i.e. 70 K, which is the transition temperature. On the same plot is shown the grains distribution $f(1/L^3)$ over their reciprocal volumes (large circles) which are proportional to the activation energies $\varepsilon_L$. Excellent agreement between those two distribution functions is evident. The energy $\langle \varepsilon_L \rangle = 6$ meV corresponds to an average grain size $(\langle L^{-3} \rangle)^{-1/3}$ of 3 nm in accordance with the theoretical estimation. In addition, in inset Fig. 8 is shown the integral distribution function $F(T) = \int_0^{kT} f(\varepsilon_L) d\varepsilon_L$, equivalent to the fraction of "switched on" constrictions at a given temperature. The larger the fraction is, the more developed is the metal cluster and the more the Hall conductivity channel is shunted.[38] As a consequence, the Hall voltage drops with temperature and such a temperature dependence has similitude with the increase of the function $F(T)$ (compare Fig. 3 & 8).

On the other hand, our noise data might also be interpreted as a direct consequence of QSE in the smallest granules.

Electrical low frequency noise in an Infinite Cluster can be modelled in term of slow charge carriers exchanges between the percolation path and isolated particles in the vicinity of the metallic network.[41] Both the carriers number fluctuations and the mobility fluctuations contribute to the electrical noise and a wide distribution of hopping energies or tunnelling distances gives rise to $1/f$ resistance fluctuations. This mechanism can be reasonably invoked to explain the slight temperature increase of the noise level above 130 K in the metallic-like regime. However, around $T_t$ and below, the random telegraph noise and the drastic increase of the *1/f* resistance fluctuations necessarily imply physical changes in the IC. Even if the power spectral density of the noise is normalised by the charge carrier density deduced from the Hall measurements, the intrinsic noise level, i.e. the Hooge's constant, increases by almost one order of magnitude as the sample is cooled through $T_t$. The enhancement of the fluctuations is not only resulting from a reduction of the effective volume of conductivity. A noisier mechanism different from the one encountered in the metallic regime of the IC is active at low temperature. The most relevant aspect of the resistance fluctuations is the random telegraph noise in the close vicinity of the temperature induced QSE transition, which provides an insight into the noise sources. From the temperature dependence of non gaussian noise, we found a 100meV activation energy of the fluctuator responsible for the RTN. This activation energy should be related by any manner to the mechanism "switching on" or "off" of the energy barrier of the conductivity cuts off. However, 100meV, which is consistent with long relaxation times, is much higher than the estimated 10meV energy barrier. So it is unlikely that 100 meV is the energy barrier responsible for temperature-induced QSE transition. Nevertheless, the 100 meV activation energy for the fluctuator is favorably compared to the Coulomb energy needed to charge isolated iron particles in the insulating matrix. Slow charge exchanges on nanoscale granules located close to the constrictions modulate the potential of the barrier transmittance which is therefore randomly lowered or increased. So the barrier conductance is alternatively switched "on" and "off" according to the local random electrical field fluctuations. In the vicinity of the transition, a little amount of barriers is active (see distributions Fig.8 ) and, consequently, affected by the local potential fluctuations. Then, singular events are expected on the IC conductivity, giving rise to current redistribution and non gaussian noise. We also have demonstrated that the number of active barriers in the IC is temperature dependent and significantly increases below 60K (Fig.8). Therefore, decreasing the temperature reinforces the amount of effective barriers modulated by the local electric field fluctuations. That unambiguously implies a drastic increase of the number of activated fluctuators with a wide energy distribution related to the iron grain size and energy barriers distribution (Fig.1 & 8). It consistently explains the temperature induced change from RTN to a large *1/f* noise we observe below 60 K. Let us mention that transition from RTN to *1/f* resistance fluctuations due to a temperature-induced increase of the number of active fluctuators with a wide energy distribution has been previously invoked in submicrometer silicon inversion layers.[42] The noise model we propose, based on random potential modulation of the barrier transmittance due to local Coulomb charges strongly support the concept of QSE transition in the IC.

Finally, we point out that an applied magnetic field on the granular Fe-SiO$_2$ films in the very close upper limit of the percolation has little influence on the electronic transport properties we describe and the low temperature electronic transition. So, we rule out any magnetic contribution to the temperature-induced QSE transition.

**Conclusion**

We examined the electronic transport properties of $Fe_{X\approx X_c}$-$SiO_2$ in the very close upper percolation limit. The temperature dependence of the conductivity, the high magnetic field Hall measurements and the electrical noise behavior strongly support a temperature-induced transition in the metallic percolating network. The transition undergoes a significant decrease of the effective volume involved in the conductivity. We argue that the electronic confinement in the critical bonds of the percolating channel are responsible for a quantization of the electronic levels. A direct consequence of such Quantum Size Effects is the formation of energy barriers which, at sufficiently low temperature, induce a cutting up of the percolation. It's worth to underline that our results provide evidence that in the very close vicinity of the percolation network, the temperature effects may induce QSE transition (resulting in a crossover from tunnelling to metal- type conduction) in the same way as a metallic fraction variation near $x_c$.

**ACKNOWLEDGMENTS**

We are grateful to Prof. B. Shklovskii and Prof. A. Lagar'kov for fruitful discussions and M. Sedova, who provides us with samples. This work was supported by the Russian Foundation for Basic Research (project nos. 99-02-16955 and 00-02-17191) and the PICS Russian-French Foundation (project no. 98-02-22037).

**Figure captions**

Fig. 1. Grain size distribution function $f(L)$ defined by TEM image on $Fe_x(SiO_2)_{1-x}$ thin films. Solid line is the Gaussian fit $f(L) \propto \exp[-(L-\langle L \rangle)^2/2\sigma_L^2]$ with the average value $\langle L \rangle=3.8$ nm and the dispersion $\sigma_L=1.9$

Fig.2. Temperature dependence of the conductivity for $Fe_x$-$SiO_2$ thins films, with $x \sim x_c$. Above 70 K, the conductivity follows a $T^{0.5}$ law. In inset is plotted the resistivity versus $T$.

Fig.3. Hall resistivity of a $Fe_{0.56}$-$SiO_2$ thin film measured in high pulsed magnetic fields for several temperatures. In inset is reported the temperature dependence of the ordinary Hall coefficient.

Fig.4. Temperature dependence of the $\Delta V_{Hp}(T)$ voltage compared with that for overall resistivity of the sample. The $\Delta V_{Hp}$ temperature dependence reveals realignment of the infinite cluster. In inset is plotted the local variation of the correlation length $L_c$ in the percolation path.

Fig.5. Temperature dependence of the power spectral density of the resistance fluctuations measured at 10 Hz. One notices the two orders of magnitude increase of the noise level around $T_t$. In inset is plotted the frequency dependence of the power spectral density of the noise in a Log-Log scale following a $1/f$ law at room temperature.

Fig.6. *a*) Time traces of the resistance fluctuations for several temperatures. A highly non-gaussian noise is observed in the vicinity of the transition temperature. *b*) Frequency dependence of the $f \times S_v(f)$ at various temperatures corresponding to the non-gaussian fluctuations. In the inset is plotted the temperature dependence of the frequency peaks

the $f\times S_v(f)$ curves. They follows an Arrhenius law from which a 100 meV activation energy is inferred.

Fig.7. Temperature dependence of the metallic conductance of the Infinite Cluster $G_{IC}(T)$ deduced from the difference between the experimental conductance and the one corresponding to the hopping regime (see text) : $G_{IC}(T) = [G_{exp}(T)–G_{hop}(T)]$. The solid lines are to guide the eyes.

Fig. 8. Distribution function $f(\varepsilon_L)$ of tunnel barriers (associated with constrictions of IC) over their activation energies (small circles) and distribution function $f(1/L^3)$ of grains over their reciprocal volumes (large circles). Solid line is the fit of experimental curve by the log-norm function $f(\varepsilon_L)\propto\exp[-\ln^2(\varepsilon_L/\varepsilon_0)/2\sigma^2]$ with $\varepsilon_0$=1.35 meV, $\sigma$=1. In inset is plotted the integral distribution function $F(T)$ of tunnel barriers showing the fraction of "switched on" constrictions at a given temperature.

Fig 1

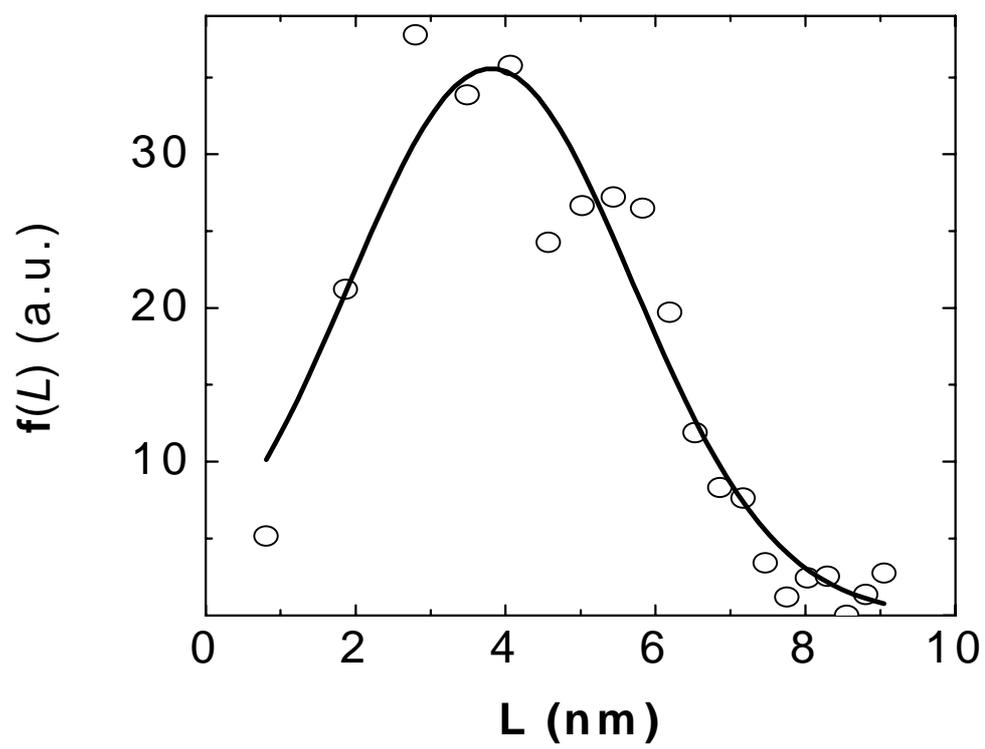

Fig. 2

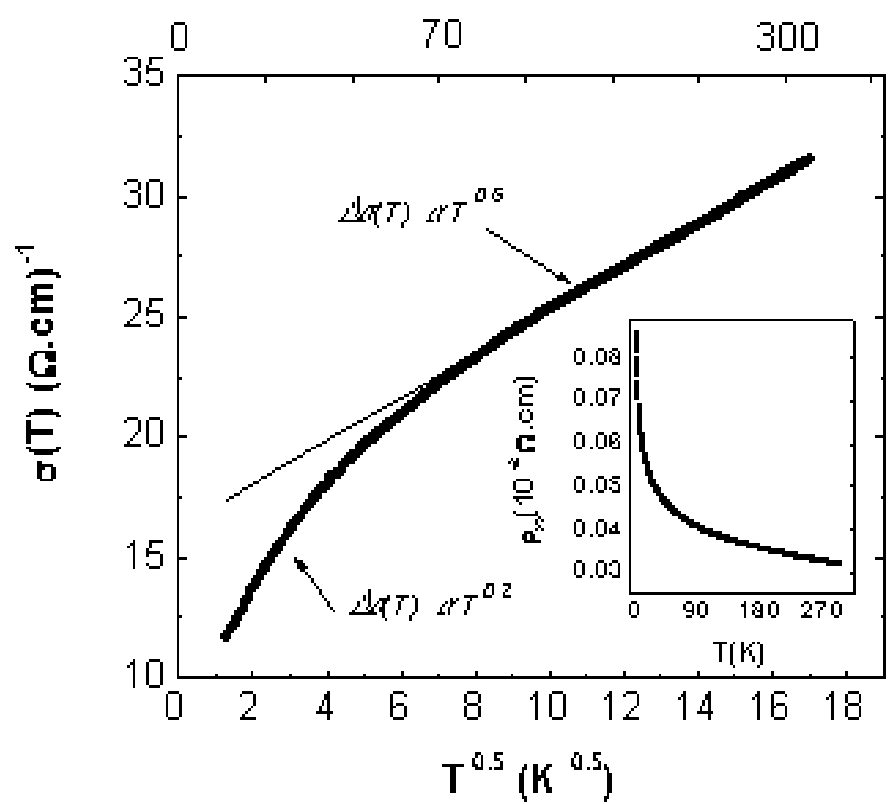

Fig. 3

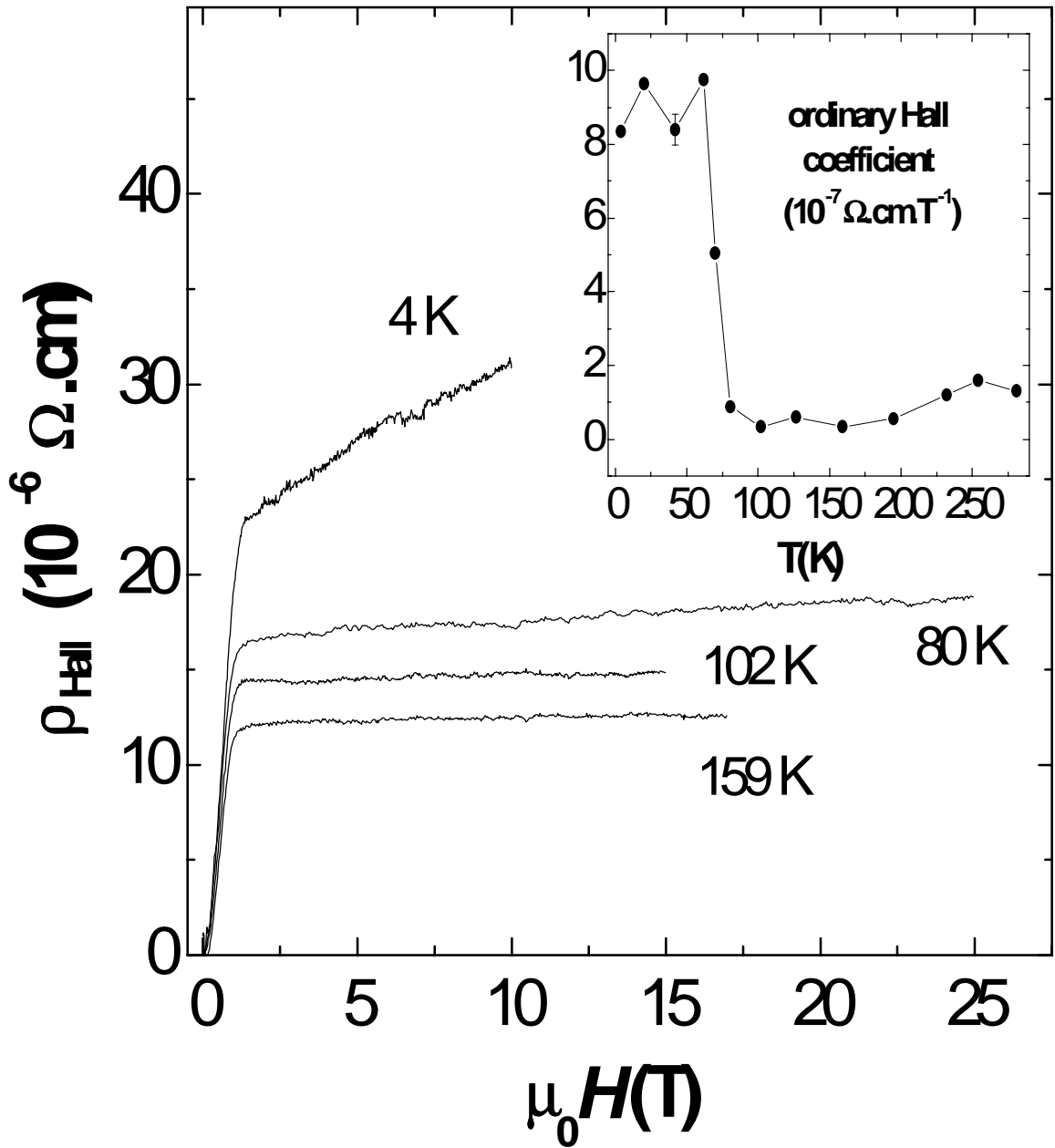

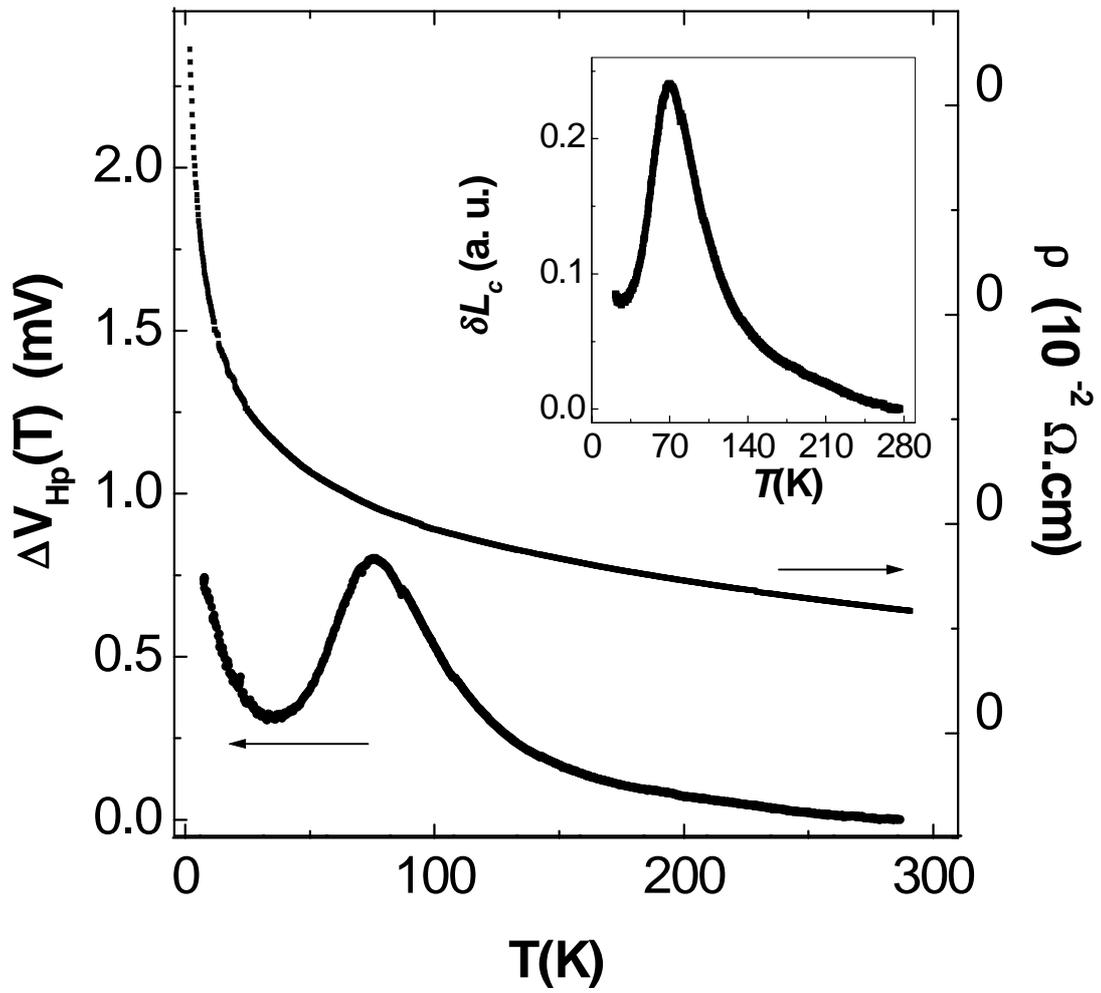

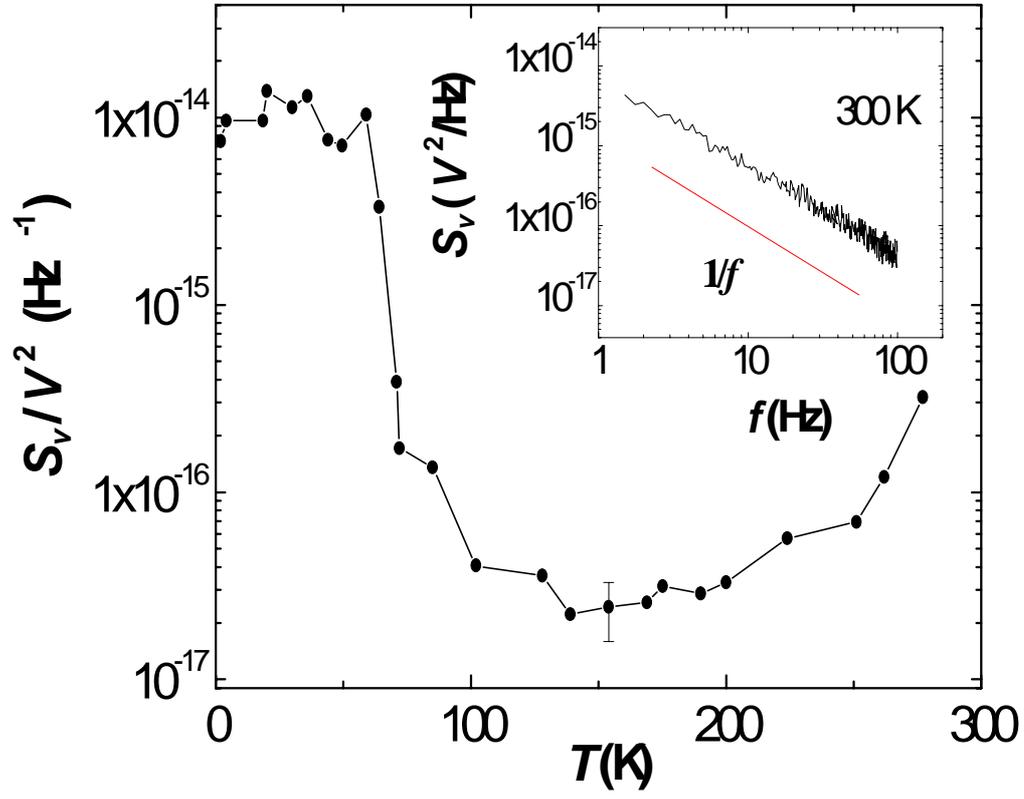

Fig. 6

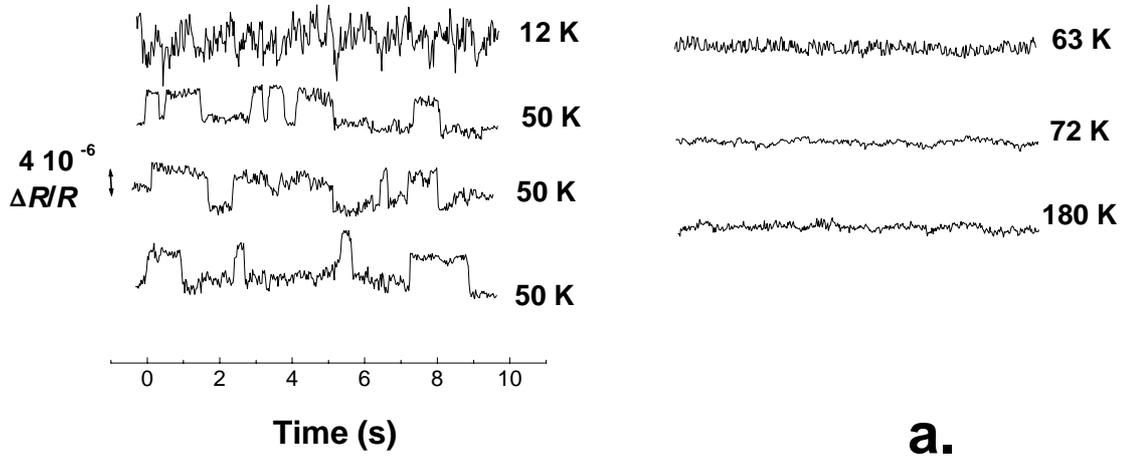

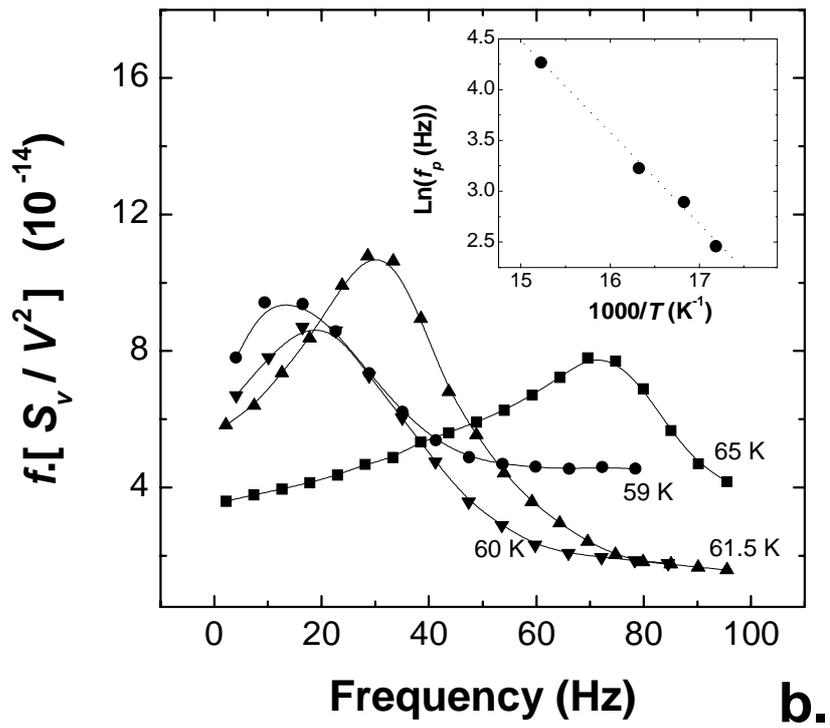

Fig. 7

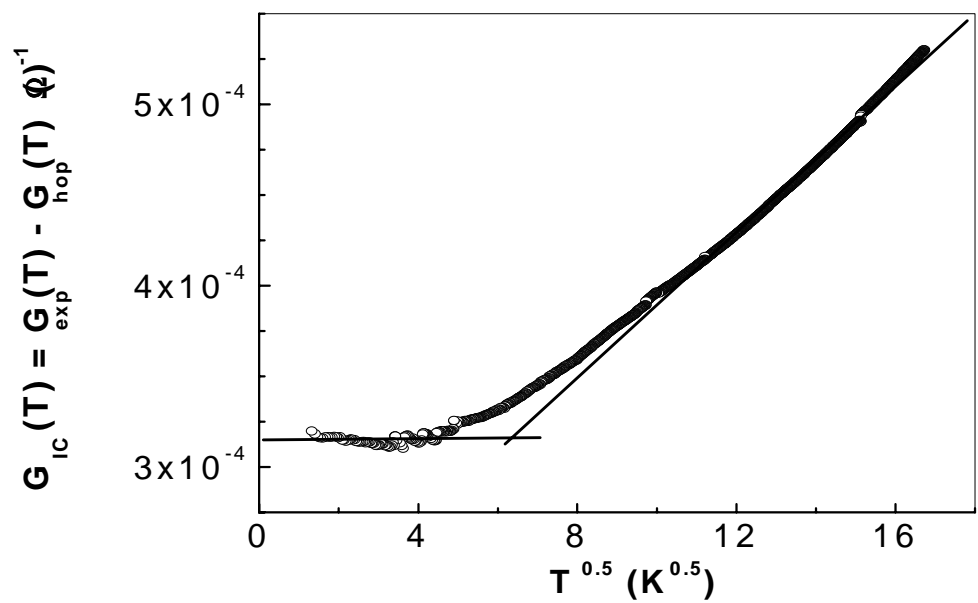

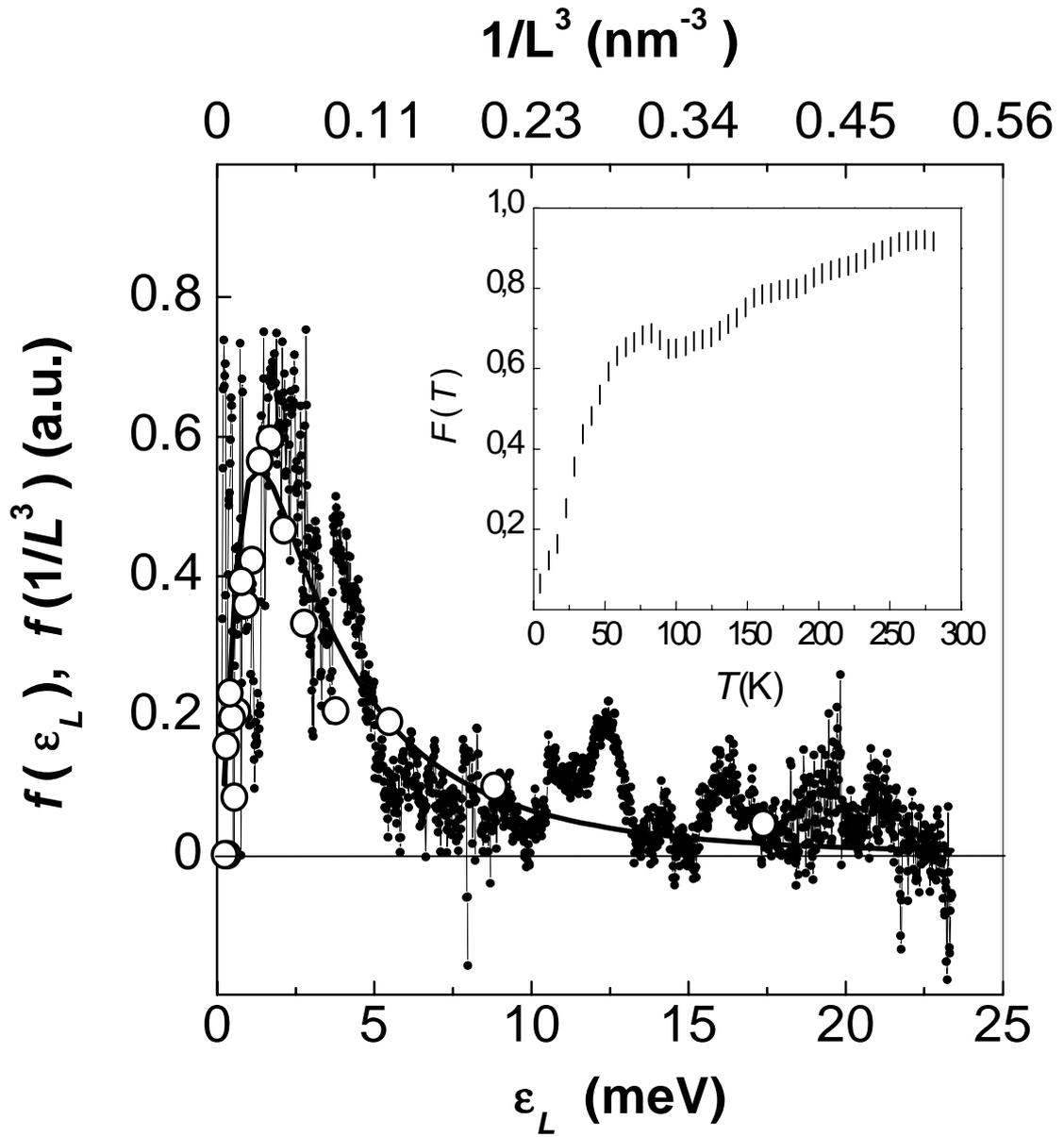